\magnification=\magstep1
\nopagenumbers
\def\b{\bigskip}
\def\s{\smallskip}
\def\c{\centerline}

\c  {\bf  Hidden assumptions in decoherence theory}\b
\c {Italo Vecchi}
\c {Bahnhofstr. 33 - 8600 Duebendorf - Switzerland}
\c { email: vecchi@weirdtech.com }
\b
\b

The present note is  devoted to a critique of some aspects of current decoherence theory. Its main point is that several common claims related  to decoherence
theory are based on questionable hidden assumptions. We refer to [1] and [2] for excellent introductions to the subject of decoherence theory and for background material.\s
We  focus on Joos´ excellent survey of decoherence theory [1], whose clarity makes its relatively easy to spot the inconsistencies in the argument.  According to the argument, given a system $ S $ in a superposition of eigenstates $ |n \rangle $ and its environment $ W $ in a state $ \Phi_o $ the pointer states are identified as those states $|\Psi(t)\rangle$ in $ W $ resulting from the interaction between  $ S $ and $ W $
$$ |n\rangle |\Phi_o\rangle \longrightarrow exp(-iH_{int})  |n\rangle |\Phi_o =: |n\rangle |\Phi_n(t)\rangle.  $$  The states $ |\Phi_n(t)\rangle $ result from the entanglement  of W with  S through the interaction Hamiltonian $ H_{int} $ and are usually referred to as the "pointer positions".  An  act of measurement on $ W $ induces a collapse of  its state vector into one of the pointer vector´s, yielding information about the state of the system $ S $. The states $ |\Phi_n(t)\rangle $ are descibed in [1] as the states of the "rest of the world". \s
The basic ambiguity underlying this description of the decoherence process may be formulated as follows. Any vector basis can be chosen as a pointer basis. The environment and any measurement device can be described using an arbitrarily chosen basis $|\Psi(t)\rangle$. The privileged pointer basis referred to by Joos is relative to an observer, as defined by a measurement operator. The measurement device or the environment do not chose a basis. The observer does. The privileged pointer basis is determined by the set of possible outcomes of a measurement act performed by an observer. It is the intervention of the observer on the measurement apparatus in the course of the measurement process that determines the pointer basis.\s
An example may clarify the underlying issue. We know that Planck's radiation law in black body theory is obtained maximising entropy on discrete energy spectra. In the black body model both  absorption and emission are continuous processes, but the entropy is maximised on discrete energy spectra. Entropy maximisation may be applied to other sets of observables too, but it will yield different results. If  the observer is associated with  continuous energy spectra then  entropy maximisation yields the Jeans-Raleigh law. Other observables yield other distribution laws. This  extends  to decoherence, so that the result of the decoherence process is seen to depend on the observer, as defined by a set of observables or, equivalently, by a measurement operator.\s
The role of the observer in the decoherence argument is indeed acknowledged in [1], as is  the fact that the superpositions in the system are not destroyed but merely cease to be identifiable by local observers. However the pointer basis is implicitly treated as an intrinsic property of the interaction between the system and its environment or a measurement device. This tacit assumption is necessary for the decay of the off-diagonal interference terms of the system's density matrix, $$
\rho_S = \sum_{n,m} c^*_m c_n |m\rangle \langle n|\longrightarrow \rho_S = \sum_{n,m} c^*_m c_n \langle \Phi_m| \Phi_n \rangle |m\rangle \langle n| $$ which is then interpreted as the vanishing of superpositions. The assumption however  leads to inconsistencies, as shown by the following analysis.
The assumption that the pointer basis is an intrinsic property of the environment would not matter if the decoherence argument was independent of the chosen pointer basis. However this is not the case. According to the argument in [1] and [2] , the decoherence process induces  the decay of the off-diagonal elements of the systems density matrix,
$$ \rho_S ´\longrightarrow \sum_n |c_n|^2 |n\rangle \langle n|  $$ which is interpreted as the emergence of a set of stable macroscopic states. The density matrix however is defined in terms of the pointer basis. Different pointer basis lead to different density matrices  for the same state vectors. It is immediate to see that the decoherence process, i.e. the decay of the off diagonal terms in the density matrix, does not commute with a change of basis. Indeed given a density matrix $ A $ , let $ C $ be a change of basis and , $ C^{-1}$ its inverse and D the operator that equates to null the off-diagonal elements. Then $$
                                     DA \neq (C^{-1} D C) A
$$
so that the result of the decoherence process depends on the pointer basis, which is selected by the observer and is independent of the underlying physical process. Indeed any two non-commuting operators induce pointer basis for which the above inequality holds. The states associated with a diagonal density matrix in one basis describe superpositions in the other basis.  
An example of different pointer basis inducing different  decoherence processes is actually considered in [2], but the authors limits themselves to pointing out the "right" pointer basis, without analysing its dependence on the observer. \s
The above  indicates that  the result of the decoherence process depends on the observer, but it also reveals that there must be a flaw in the decoherence argument. The claim that interaction with the environment induces the diagonalisation of the system's density matrix must be wrong, since the diagonalisation process depends on the chosen basis, which is not an intrinsic property of the environment but of the observer. Indeed if one examines  the argument leading to the diagonalisation of the system's density matrix, one discovers that it is based on the unphysical no-recoil assumption on the scattering process ([1]), which serves the sole purpose of preventing the environment from eroding the diagonal elements of the system's density matrix. Under the no-recoil assumption interaction with the environment action can only deplete the off-diagonal elements of the system's density matrix. The no-recoil assumption forces the density matrix into a very singular form, where the off-diagonal terms converge rapidly to zero, while the diagonal termss remains intact . Applying the no-recoil assumption to a different basis however leads to a diagonal matrix  describing a different physical state and which is not diagonal under a change of basis, as shown above. \s
The only possibility to preserve the consistency of the decoherence argument is to acknowledge that the decoherence process induces the decay of all matrix elements, since indeed if $ DA = 0 $ for any matrix $ A $ then  the equality $
                                     DA = (C^{-1} D C) A
$ holds for  for any change of basis $ C $. In that case the decoherence process describes the observer's loss of information, not only on superpositions, but on the state of the system. The special status of superpositions is indeed spurious, since it depends on the measurement operator being considered, i.e. on the observer. The singling out of superpositions for special destructive treatment  appears as an  anthropomorphic artefact, based on unphysical assumptions. \s
Certain results of decoherence theory's preserve their validity in the light of the above criticism. Indeed Zurek´s "predictability sieve" ([3]) may be reformulated as the claim that if a system's state can be tracked  by an observer, it will behave as expected , so that decoherence reflects the inability of an observer subject to the second principle of thermodynamics to keep track of the system's interaction with the  environment.\s
The pointer basis is a privileged reference system. The belief that Nature does not provide privileged reference systems, unless nudged into doing so by  anthropomorphic  assumptions, provided the motivation for this note. \b

 \c{ \bf References} \b

[1] A. Joos "Decoherence Through Interaction with the Environment" in "Decoherence and the Appearance of a Classical World in Quantun Theory" D.Giulini et al. ed,, Spinger, Berlin-Heidelberg-New York 1999.\s
[2] J.P. Paz , W.H. Zurek "Quantum limits of decoherence: Environment induced superselection of energy eigenstates" Phys.Rev.Lett. 82 (1999) 5181-5185.\s
[3] Zurek W.H. "Preferred Observable of Predictability, Classicality and the Environment Induced Decoherence" Progr. Theor. Phys. 89, 281-312.

\end